\documentclass[conference]{IEEEtran}
\IEEEoverridecommandlockouts
\usepackage{cite}
\usepackage{amsmath,amssymb,amsfonts}
\usepackage{algorithmic}
\usepackage{graphicx}
\usepackage{booktabs}
\usepackage{multirow}
\usepackage{textcomp}
\usepackage{xcolor}
\usepackage{hyperref}
\def\BibTeX{{\rm B\kern-.05em{\sc i\kern-.025em b}\kern-.08em
    T\kern-.1667em\lower.7ex\hbox{E}\kern-.125emX}}
\begin{document}

\title{ProPS: Prompted Profile Synthesis for Natural Language-Conditioned Speaker Embedding Distributions}


\author{\IEEEauthorblockN{Thomas Thebaud\IEEEauthorrefmark{1}\IEEEauthorrefmark{2}\thanks{Work done at the Johns Hopkins University.}, Junhyeok Lee\IEEEauthorrefmark{1}, Laureano Moro-Velazquez\IEEEauthorrefmark{1}\IEEEauthorrefmark{3}, Jesus Villalba Lopez\IEEEauthorrefmark{1}, Najim Dehak\IEEEauthorrefmark{1}}
\IEEEauthorblockA{\IEEEauthorrefmark{1}\textit{Department of Electrical and Computer Engineering, Johns Hopkins University}, Baltimore, MD, USA}
\IEEEauthorblockA{\IEEEauthorrefmark{2}\textit{AMIAD}, Palaiseau, France. 
Email: \textit{thomas.thebaud@polytechnique.edu}}
\IEEEauthorblockA{\IEEEauthorrefmark{3}\textit{Department of Signal Theory and Communications, Universidad Carlos III de Madrid (UC3M)}, Leganes, Spain}
}

\maketitle

\begin{abstract}
Speaker embeddings, or x-vectors, are widely used to represent speaker identity and speaker-related attributes, but existing embedding extractors are typically descriptive rather than generative: they map an observed speech segment to an x-vector, which is then used for downstream applications. 
We introduce ProPS, Prompted Profile Synthesis, a framework for generating distributions of speaker embeddings conditioned on natural language prompts such as "a thirties male speaker with an Indian accent".
ProPS converts human-written profile descriptions into sentence embeddings and uses a mixture density network trained on a large-scale dataset to predict a Gaussian mixture model in the x-vector space. 
The model is trained by maximizing the likelihood that real speaker embeddings match the requested profile, and its generated distributions are evaluated by negative log-likelihood on held-out x-vectors and by attribute classification accuracies on sampled synthetic x-vectors. 
Experiments show that ProPS produces profile-conditioned distributions and generates x-vectors that preserve requested speaker attributes such as age, gender, accent, and prosodic characteristics. 
This design enables controllable speaker-profile synthesis for speech generation systems like Text-To-Speech (TTS) or Voice Conversion (VC) while anchoring generated distributions in observed speaker-embedding structure.

\end{abstract}

\begin{IEEEkeywords}
X-vectors, mixture density networks, controllable generation, synthetic speakers, natural language prompts
\end{IEEEkeywords}

\section{Introduction}
Human voices carry information beyond linguistic content, including speaker identity, demographic background, accent, speaking style, prosody, and recording conditions. A large body of work has studied how to extract, represent, compare, and profile speaker identities from speech signals. As speech technologies become increasingly generative and controllable, modeling not only individual identities but also broader speaker profiles becomes an important research problem.

Neural speaker embeddings, commonly referred to as x-vectors~\cite{snyder2018x}, are now the dominant representation for speaker information. They map speech segments to fixed-dimensional vectors that capture speaker-related characteristics while abstracting away from the raw waveform. X-vectors and related embeddings are widely used in speaker verification~\cite{snyder2018x}, diarization~\cite{palka2026vbx}, voice conversion~\cite{qian2019autovc}, anonymization~\cite{srivastava2022privacy}, text-to-speech~\cite{cooper2020zero}, speaker profiling~\cite{yang2025demographic}, and healthcare applications~\cite{favaro2023interpretable}. In generative systems, they often serve as compact conditioning variables for synthesizing, converting, or adapting speech toward a desired voice identity.

When reference audio is available, an enrollment utterance can be used to extract the desired x-vector. Without such audio, controllable generation of new speaker identities remains limited. Existing approaches often sample synthetic embeddings from distributions fitted to training x-vectors, such as Gaussian mixture models or PLDA-style priors. While these methods can generate new points in speaker space, their controllability is constrained by the labels, structure, and biases of the data used to estimate the distribution.

We propose ProPS, Prompted Profile Synthesis, a framework for natural language-conditioned generation of x-vector distributions. 
Instead of sampling x-vectors from an unconditional or weakly conditioned speaker prior, ProPS takes a text prompt describing a speaker profile (such as "a young female speaker with an Indian accent and a fast pace") and generates a Gaussian mixture model in the x-vector space. 
The prompt is embedded using a sentence encoder, and a mixture density network maps this representation to a profile-conditioned distribution over speaker embeddings. In this way, ProPS treats natural language as a flexible control interface for speaker-profile synthesis, enabling users to describe desired speaker characteristics directly rather than selecting from a fixed set of predefined classes.

Our contributions are as follows:
\begin{enumerate}
\item We introduce the task of natural language-conditioned speaker-distribution synthesis, where a free-form textual description is used to generate a full distribution over x-vectors rather than sampling a single speaker embedding.
\item We propose \textbf{ProPS}, a prompt-conditioned mixture density network that maps sentence-level text embeddings to Gaussian mixture models in the x-vector space.\footnote{\url{https://huggingface.co/ThomasThebaud/PROPS_K16}}
\item We leverage the large-scale \textbf{CapSpeech}~\cite{wang2025capspeech} speaker-profile data to learn associations between natural language descriptions, speaker attributes, prosodic characteristics, and observed x-vector distributions.
\item We evaluate the generated x-vector distributions using held-out negative log-likelihood and downstream attribute classifiers, measuring both distributional fit and controllability with respect to gender, age, accent, and prosodic characteristics.
\end{enumerate}

\section{Related Work}
\subsection{Speaker Embeddings}
Speaker embeddings provide compact fixed-dimensional representations of speaker identity and related characteristics. Early systems relied on i-vectors, which modeled speakers in a low-dimensional latent space derived from Gaussian mixture model supervectors and became a standard approach for speaker verification~\cite{dehak2010front}. Deep neural speaker embeddings, or x-vectors, later improved robustness and verification performance by training neural networks to discriminate among speakers~\cite{snyder2018x}. More recent architectures such as ECAPA-TDNN further improved speaker representation learning through channel attention, propagation, and multi-layer feature aggregation, producing embeddings widely used in speaker recognition, voice conversion, speech synthesis, and other speaker-conditioned generation tasks~\cite{desplanques2020ecapa}.

\subsection{Mixture Density Networks}
Mixture Density Networks (MDNs) were introduced by Bishop~\cite{bishop1994mixture} to model multimodal conditional distributions by combining neural networks with Gaussian mixture models. Originally motivated by inverse problems such as robot inverse kinematics, MDNs address situations where multiple valid outputs may correspond to the same input, a setting in which conventional regression networks tend to predict only the conditional mean. This property makes MDNs particularly well suited to speaker-profile synthesis, where a single textual description may correspond to many plausible speaker identities.

\subsection{Natural Language Representations}
Transformer architectures have reshaped natural language processing by enabling large-scale self-supervised learning of contextualized token representations through self-attention~\cite{vaswani2017attention}. BERT demonstrated that bidirectional transformer pretraining on large text corpora could produce transferable representations for many downstream NLP tasks~\cite{devlin2019bert}. Sentence-BERT (SBERT) extended this paradigm to sentence-level representation learning by using a Siamese transformer architecture optimized for semantic similarity, producing fixed-dimensional embeddings that capture sentence meaning and can be efficiently compared with vector operations~\cite{reimers-2019-sentence-bert}. These properties make SBERT embeddings well suited for natural language conditioning in downstream generative models.

\subsection{Generating X-vectors from natural language prompts}
PromptSpeaker~\cite{zhang2023promptspeaker} is closely related to our work, as it also addresses the generation of speaker embeddings from natural language descriptions. In this approach, a prompt encoder maps each textual description to the parameters of a Gaussian distribution in a semantic latent space, and semantic embeddings sampled from this distribution are then transformed into x-vectors through an invertible Glow model~\cite{kingma2018glow}. This formulation enables multiple speaker embeddings to be generated from the same prompt, but the diversity of the generated x-vectors is controlled indirectly through sampling in the semantic space and is constrained by a single Gaussian distribution. In contrast, ProPS directly models a prompt-conditioned Gaussian mixture distribution in x-vector space, providing explicit control over the generated speaker-embedding distribution. Such control is important for downstream voice conversion and text-to-speech applications, where generated voices should not only match the requested profile but also remain diverse and distinct from existing speakers.

\begin{figure*}[t]
    \centering
    \vspace{-5mm}
    \includegraphics[width=0.95\linewidth]{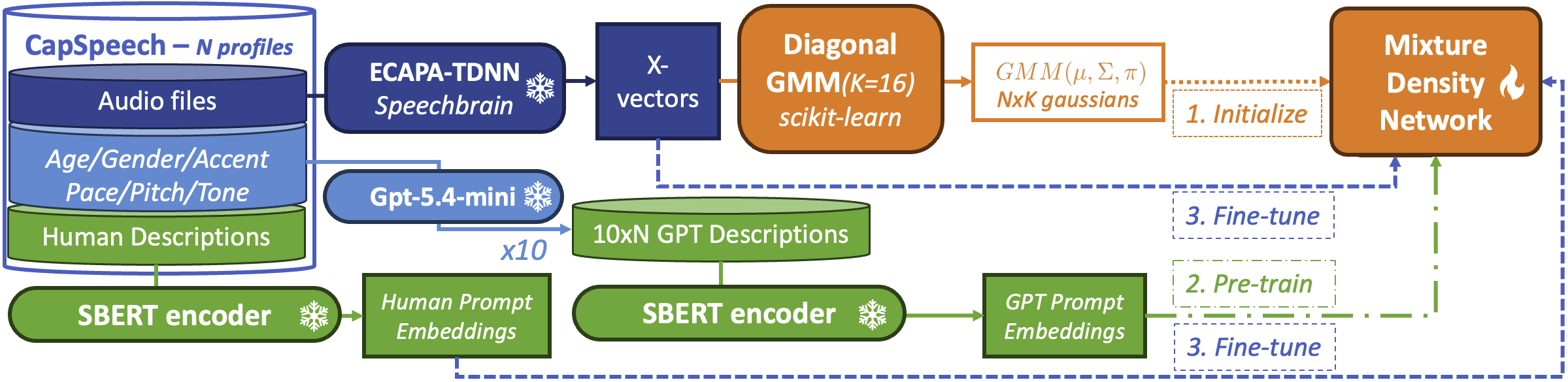}
    \caption{Overview of the training pipeline of ProPS.}
    \label{fig:overview}
    \vspace{-5mm}
\end{figure*}

\section{Method}
\label{sec:method}

\subsection{Problem Formulation}
Let $p$ denote a natural language prompt describing a speaker profile, and let $e = f_{\mathrm{text}}(p) \in \mathbb{R}^{d_t}$ be the corresponding text embedding. Let $x \in \mathbb{R}^{D}$ denote an ECAPA-TDNN speaker embedding extracted from a speech segment. The goal is to learn a conditional density:
\begin{equation}
q_{\theta}(x \mid p) = q_{\theta}(x \mid e)
\end{equation}
where $q_{\theta}$ is a GMM predicted by a neural network. 
For a $K$-component GMM:
\begin{equation}
q_{\theta}(x \mid e) = \sum_{k=1}^{K} \pi_k(e)\mathcal{N}\left(x;\mu_k(e),\mathrm{diag}(\sigma_k^2(e))\right)
\label{eq:gmm}
\end{equation}
where $\pi_k(e)$ are non-negative mixture weights that sum to one, $\mu_k(e)$ are component means, and $\sigma_k(e)$ are diagonal standard deviations.

Given a training set of profile descriptions and matching real x-vectors, the model is trained by minimizing negative log likelihood:
\begin{equation}
\mathcal{L}_{\mathrm{NLL}}(\theta)
= - \frac{1}{N} \sum_{i=1}^{N} \log q_{\theta}(x_i \mid e_i).
\label{eq:nll}
\end{equation}
Optional regularization terms encourage stable mixture weights, useful entropy, and bounded variance, but the primary objective is likelihood of real x-vectors under the prompt-conditioned distribution.

\subsection{Dataset and Profile Description Protocol}
\subsubsection{Dataset}
We train and evaluate ProPS on CapSpeech~\cite{wang2025capspeech}, a large-scale speaker-profile dataset assembled from  GigaSpeech~\cite{chen2021gigaspeech}, the English subsets of CommonVoice~\cite{ardila2019common}, MLS~\cite{pratap2020mls}, and Emilia~\cite{he2024emilia}. 
CapSpeech associates speech segments with structured speaker-profile metadata, including demographic attributes such as age, gender, and accent, as well as prosodic descriptors such as pitch, pace, and tone. 
Each audio segment is also associated with a human-written description.

\subsubsection{Profile characteristics}
The full dataset contains more than 31k hours of speech and 9372k segments (of which 9275k are in the train split, 48k are in the development, and 48k are in the test split). 
We select only profiles with at least 100 corresponding segments, yielding 942 profiles spanning gender, age, accent, pitch, pace, and tone attributes.
They include:
\begin{itemize}
    \item \textbf{Pitch}: very low, low, slightly low, moderate, slightly high, high, very high
    \item \textbf{Age}: teenager, young adult, middle-aged adult, elderly
    \item \textbf{Gender}: female, male
    \item \textbf{Pace}: slowly, slightly slow, moderate speed, slightly fast, fast, very fast
    \item \textbf{Tone}: very monotone, monotone, slightly expressive and animated, expressive and animated, very expressive and animated
    \item \textbf{Accent}: American, Australian, Belgian, Brazilian, British, Canadian, Cantonese, Czech, English, Filipino, French, German, Hungarian, Indian, Irish, Italian, Japanese, New Zealand, Norwegian, Russian, Scottish, Slovenian, Welsh
\end{itemize}

\subsubsection{Generated descriptions}
To connect structured profiles with free-form natural language, we use both generated descriptions and human-written descriptions.
We generate textual descriptions for each unique profile by submitting the corresponding attribute combination to ChatGPT (model \texttt{gpt-5.4-mini-2026-03-17})~\cite{openai2024gpt4technicalreport} and requesting 10 concise yet distinct natural-language descriptions of the same speaker profile.
For example, a structured profile containing \texttt{gender=male}, \texttt{age=thirties}, \texttt{accent=Indian}, and \texttt{pace=measured} may be rendered as \texttt{"a thirties male with an Indian accent speaking with a measured tone"}. 
The 10 GPT-generated descriptions for each profile will later be referred to as descriptions 1 through 10.

\subsubsection{Embedding extraction}
Prior to training, the semantic and speaker embeddings are extracted using their corresponding encoders.
Each natural language description, both human-written and GPT-generated, is encoded with a pre-trained sentence BERT~\cite{reimers-2019-sentence-bert} (SBERT model \texttt{
all-MiniLM-L6-v2}), yielding 384-dimensional semantic embeddings. 
Each spoken utterance is encoded using a pretrained ECAPA-TDNN model~\cite{desplanques2020ecapa} from the speechbrain toolkit~\cite{speechbrain} with dimensionality $D=192$.

\subsection{Proposed Architecture}
Assume that the training set contains $N$ unique speaker profiles. For each profile, we precompute a diagonal-covariance GMM with $K$ Gaussian components from the x-vectors associated with that profile. The collection of these profile-level GMMs defines a shared component bank with $N \times K$ Gaussian components. Each component has a mean vector $\mu \in\mathbb{R}^{D}$ and a diagonal standard deviation vector $\sigma \in \mathbb{R}^{D}$, where $D=192$ is the x-vector dimensionality.

Given a natural language prompt, ProPS first computes an SBERT embedding $e \in \mathbb{R}^{384}$. This embedding is fed into a three-layer multilayer perceptron with hidden dimensions 1024, 2048, and 1024. The MLP outputs $N \times K$ logits, which are transformed with a softmax to obtain mixture weights over the full component bank $
\pi(e) = \mathrm{softmax}\left(\mathrm{MLP}_{\theta}(e)\right)$ with $\pi(e) \in \mathbb{R}^{N K}$.
The Gaussian means and diagonal standard deviations of the MDN are trainable parameters, initialized from the precomputed profile GMMs: $\mu \in \mathbb{R}^{N K \times D}$ and $\sigma \in \mathbb{R}^{N K \times D}$.
The resulting conditional density is therefore
\begin{equation}
q_{\theta}(x \mid e)
=
\sum_{j=1}^{N K}
\pi_j(e)
\mathcal{N}
\left(
x;
\mu_j,
\mathrm{diag}(\sigma_j^2)
\right).
\label{eq:ProPS_architecture}
\end{equation}
In this architecture, the text encoder and MLP learn how a natural language prompt should select and combine components from the bank of observed speaker-profile distributions. The component means and variances provide an initialization grounded in real x-vector statistics, while the learned mixture weights provide prompt-level controllability.

\subsection{Three-Stage Training Procedure}
The training proceeds in three stages:

\subsubsection{Initialization: Profile-wise GMMs}
First, for each of the $N$ profiles in the training set, we collect all associated x-vectors and fit a diagonal-covariance GMM using scikit-learn~\cite{kramer2016scikit}. Each profile GMM contains $K$ Gaussian components and estimates profile-specific mixture weights, means, and diagonal standard deviations directly from real x-vectors. The resulting $N$ profile GMMs define a shared component bank with $N \times K$ Gaussian components. Their means and diagonal standard deviations are used to initialize the trainable MDN parameters $\mu$ and $\sigma$.

\subsubsection{Pre-training: Setting the Mixture Weights}
Second, we train the MLP to predict mixture weights over the full $N \times K$ component bank from semantic prompt embeddings. For each profile, GPT-generated descriptions 3 through 10 are used as training prompts. The SBERT embedding $e$ of each description is passed through the MLP, which outputs a distribution $\pi(e)$ over all $N \times K$ Gaussian components. The target distribution is constructed from the precomputed GMM of the corresponding profile: the $K$ components associated with that profile receive their precomputed mixture weights, while all other components receive zero weight. Validation is performed using GPT-generated description 2 on the development set. This stage teaches $\mathrm{MLP}_{\theta}(e)$ to map paraphrased profile descriptions to the appropriate region of the component bank.

\subsubsection{Fine-tuning: End-to-End Training with Human Descriptions}
Third, the MDN is fine-tuned end-to-end using human-written descriptions and their associated x-vectors. Each training example consists of a natural language speaker description and a corresponding target x-vector. The description is encoded with SBERT, passed through the MLP to produce prompt-conditioned mixture weights, and the likelihood of the target x-vector is computed under the resulting GMM. The model is optimized by minimizing the negative log-likelihood loss in Eq.~\ref{eq:nll}. During fine-tuning, the model updates both the prompt-conditioned mixture weights and the Gaussian parameters to improve the fit between generated speaker-profile distributions and observed x-vectors. This final stage aligns the model with real human descriptions and encourages the generated distributions to preserve both demographic and prosodic speaker characteristics.

\subsection{Inference}
At inference time, ProPS takes a natural-language prompt describing the desired speaker profile. 
The prompt is encoded with SBERT and passed through the trained MLP to obtain mixture weights over the $N \times K$ Gaussian component bank. 
Together with the fine-tuned means and diagonal standard deviations, these weights define a prompt-conditioned GMM in the x-vector space. 
Synthetic speaker embeddings are then generated by sampling from this GMM. 
The generated x-vectors can be used directly for downstream analysis or as conditioning representations for speaker-dependent speech generation systems.

\section{Experimental Setup}

\subsection{Training Configuration}
All experiments are run either on CPU nodes or on NVIDIA GTX 1080 Ti GPUs. 
The ProPS MDN is initialized from the profile-level GMMs described in Section~\ref{sec:method}. 
Unless otherwise specified, these GMMs are computed on the CapSpeech training set with $K=16$ diagonal Gaussian components per profile. 
The resulting means and diagonal standard deviations initialize the Gaussian component bank of the MDN, while the MLP is trained to predict the corresponding mixture weights from natural language prompt embeddings.

Training proceeds in two neural optimization stages. 
First, the MDN is pretrained for 100 epochs using the GPT-generated profile descriptions and a cross-entropy loss on the precomputed GMM mixture weights. 
This stage uses a learning rate of $10^{-4}$ and is run until the development loss converges. 
Second, the initialized MDN is fine-tuned for 100 epochs on the CapSpeech training set using human speaker descriptions and their associated x-vectors. 
This fine-tuning stage uses a negative log-likelihood loss and a learning rate of $10^{-5}$. Training is stopped after convergence of the NLL on the development set. The training and evaluation code is available on GitHub\footnote{\url{https://github.com/thomasthebaud/PROPS}}.

\subsection{Choosing the Number of Gaussian Components}
The number of Gaussian components $K$ controls the expressiveness of the profile-level GMMs used to initialize ProPS. 
A small value of $K$ may underfit the diversity of x-vectors within a profile, while a large value of $K$ may overfit individual training samples and reduce the number of profiles for which a valid GMM can be estimated. 
To select an appropriate value, we fit independent profile-level GMMs on the CapSpeech training set with $K \in \{2^N | N\in [\![0,8]\!]\}$.
For each value of $K$, we compute the negative log-likelihood of held-out development and test x-vectors under the corresponding profile-conditioned GMMs.

Because we use diagonal covariance matrices, each Gaussian component must be estimated from a sufficient number of x-vectors. We therefore require at least two x-vectors per Gaussian component, meaning that a profile is considered valid for a given $K$ only if it contains at least $2K$ training x-vectors. As $K$ increases, the number of valid profiles decreases. We report both the development and test NLL and the number of valid training profiles for each value of $K$.

\subsection{Impact of Each Training Stage}
We performed an ablation study to quantify the contribution of each stage of the ProPS pipeline. 
Specifically, we compare the four steps of model training using development and test negative log-likelihood:
\begin{enumerate}
    \item \textbf{Random GMMs}: the randomly initialized GMMs with $K=16$ and diagonal covariances. This model is independent of the prompt.
    \item \textbf{Precomputed GMMs}: the GMMs trained on each individual profile. Each GMM is scored against the vectors matching its profile, but no description is used.
    \item \textbf{Pretrained MDN}: the pretrained network is evaluated both on the human-written description from CapSpeech and the unseen GPT descriptions (index 0 for the test and 1 for the dev).
    \item \textbf{Fine-tuned MDN}: the final system is evaluated on the same prompts as the \textit{Pretrained} system, but it has also seen human prompts during finetuning.
\end{enumerate}
We expect the random model to obtain the worst NLL, since it is not anchored in the observed x-vector distribution. 
The precomputed GMMs should provide a strong likelihood baseline. 
The pretrained MDN should slightly degrade performance, as it learns to match natural-language prompts to the correct GMMs.
Finally, the fine-tuned MDN should improve the NLL on the human descriptions.

\subsection{Evaluation of Speaker-Characteristic Accuracy}
We evaluate whether the generated x-vectors preserve the requested speaker characteristics using downstream attribute classifiers. For each characteristic, such as gender, age, accent, pitch, pace, or tone, we construct minimal natural language prompts that specify only that characteristic. For example, for the gender characteristic, we use prompts such as "a male speaker" and "a female speaker." Each prompt is passed through the trained ProPS model to generate a characteristic-conditioned GMM.

To measure whether the generated distribution encodes the requested attribute, we train a support vector machine classifier on real development-set x-vectors. The classifier is trained separately for each characteristic, for example, to distinguish male from female speakers for gender classification. 
For independent classes (gender and accent) we use a SVC, but for classes that present an order of labels (age, pace, pitch, tone), we attribute each class to a number and train a SVR model. 
SVR predictions are rounded/clipped to the nearest class before computing accuracy.
We first evaluate the classifier on real test-set x-vectors to estimate the accuracy achievable on real held-out speech embeddings.
We then sample 10,000 synthetic x-vectors from each generated GMM and evaluate the same classifier on these generated samples.

\section{Results}

\subsection{Effect of the Number of Gaussian Components}

\begin{figure}[ht]
\centering
\vspace{-3mm}
\includegraphics[width=\linewidth]{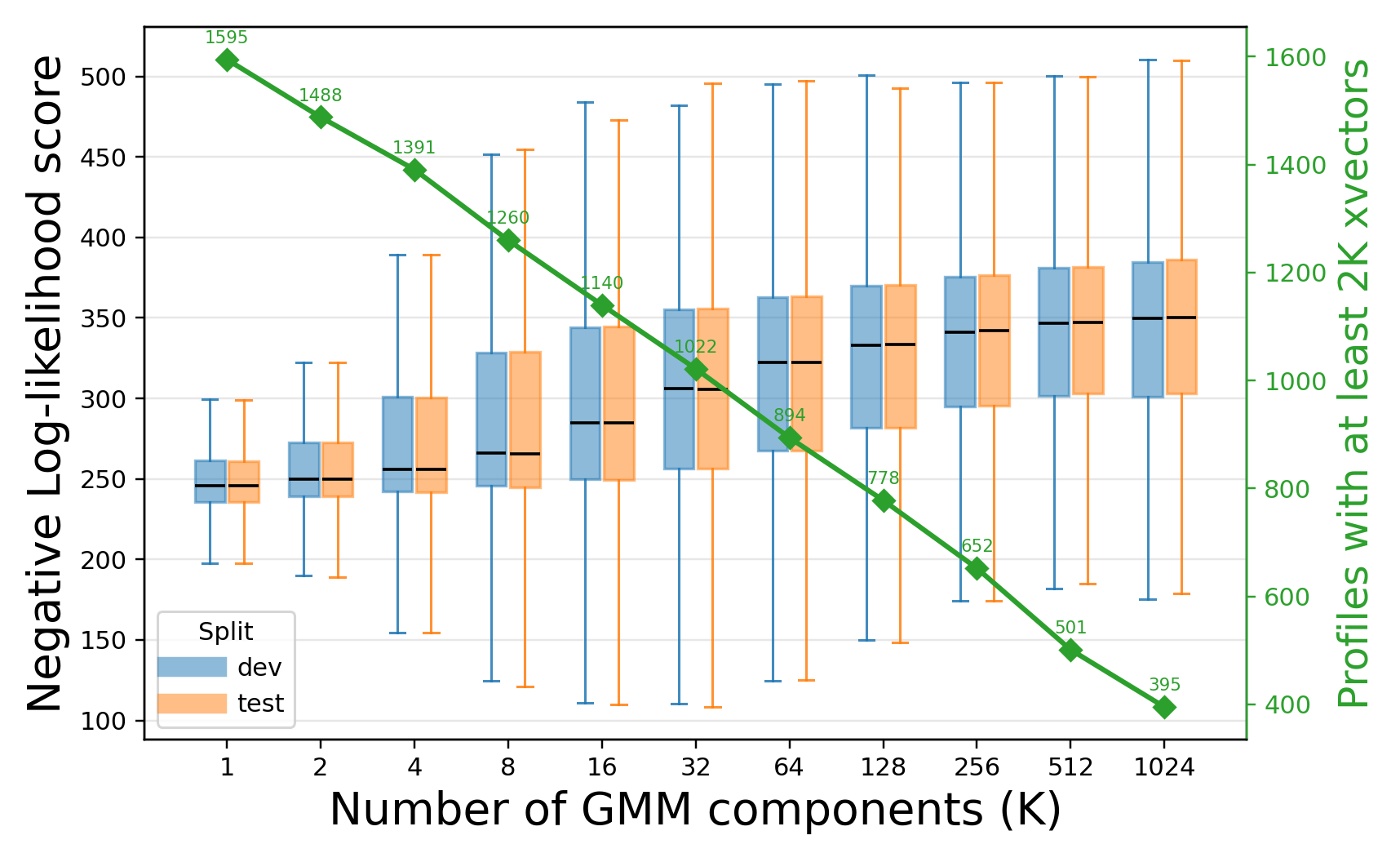}
\vspace{-5mm}
\caption{Log-likelihood of the dev and test sets as a function of the number of Gaussian components per profile. Models are compared for $1\leq K \leq 1024$. For each $K$, the number of profiles with at least $2K$ segments is also shown in green.}
\label{fig:nll_k}
\vspace{-5mm}
\end{figure}

We first study the effect of the number of Gaussian components in the profile-level GMMs. We compare values $K \in \{2^n : n=0,\ldots,8\}$ and evaluate train, development, and test negative log-likelihood. As shown in Fig.~\ref{fig:nll_k}, increasing $K$ improves the fit to held-out x-vectors but reduces the number of profiles with enough data for training. Since the NLL stabilizes for $K \geq 16$, we use $K=16$ in the remaining experiments to balance likelihood and profile coverage. Models trained with other values of $K$ did not show meaningful differences in the following experiments and are therefore omitted.

\subsection{Effect of the Fine-tuning}

We evaluate the contribution of the different initialization and training stages by comparing log-likelihood scores on development and test x-vectors. Table~\ref{tab:finetuning} reports the LL obtained when the conditioning prompt is either GPT-generated or human-written. 
For the test set, we use the held-out GPT-generated description number 1 and the human-written CapSpeech descriptions. 
For the development set, we use GPT-generated description 2 and the human-written CapSpeech descriptions. 
A Welch's T-test is performed using the SciPy toolkit~\cite{virtanen2020scipy} between consecutive lines of results to verify whether the differences are significant.

\begin{table}[t]
    \centering
    \caption{LL scores$\uparrow$ between the real x-vectors and the distributions computed from GPT-generated prompts and human-written prompts, by different models. Highest results in \textbf{bold}, \underline{Underlined} results are significantly different from the line above (here p-value$\leq10^{-3}$)}
\resizebox{\linewidth}{!}{%
\begin{tabular}{l|cc|cc}
\toprule
\multirow{2}{*}{Model} & \multicolumn{2}{c|}{Test} & \multicolumn{2}{c}{Dev} \\
 & GPT 1 & Human & GPT 2 & Human \\
\midrule
Precomputed GMMs & 303.8 & 303.1 & 303.6 & 302.9 \\
Pretrained MDN & 304.2 & 303.5 & \textbf{304.0} & 303.4 \\
Fine-tuned MDN & \textbf{\underline{305.3}} & \textbf{\underline{309.1}} & 303.7 & \textbf{\underline{308.9}} \\
\bottomrule
\end{tabular}
}
\label{tab:finetuning}
\vspace{-5mm}
\end{table}

The precomputed GMMs already provide a strong likelihood baseline, as expected given the existing literature on modeling speaker behavior with GMMs~\cite{mclaughlin1999study, zheng2004text, al2017comparison}.
However, this baseline does not include any controllability, which will be added in the next training stage.

The pretrained MDN performs similarly to the GPT descriptions, indicating that the pretraining efficiently matched the artificial prompts to the correct GMMs. 
The human-written descriptions show a non-significantly lower likelihood, suggesting that the GPT descriptions generated from the characteristics were semantically close to the human-written descriptions.

Fine-tuning brought a significant improvement in the human descriptions and a non-significant drop in the GPT-generated descriptions of the test set, compared with the dev set (p-value=0.21).

\subsection{Qualitative Analysis of Generated X-Vector Distributions}

\begin{figure}[t]
    \centering
    \vspace{-5mm}
    \includegraphics[width=\linewidth]{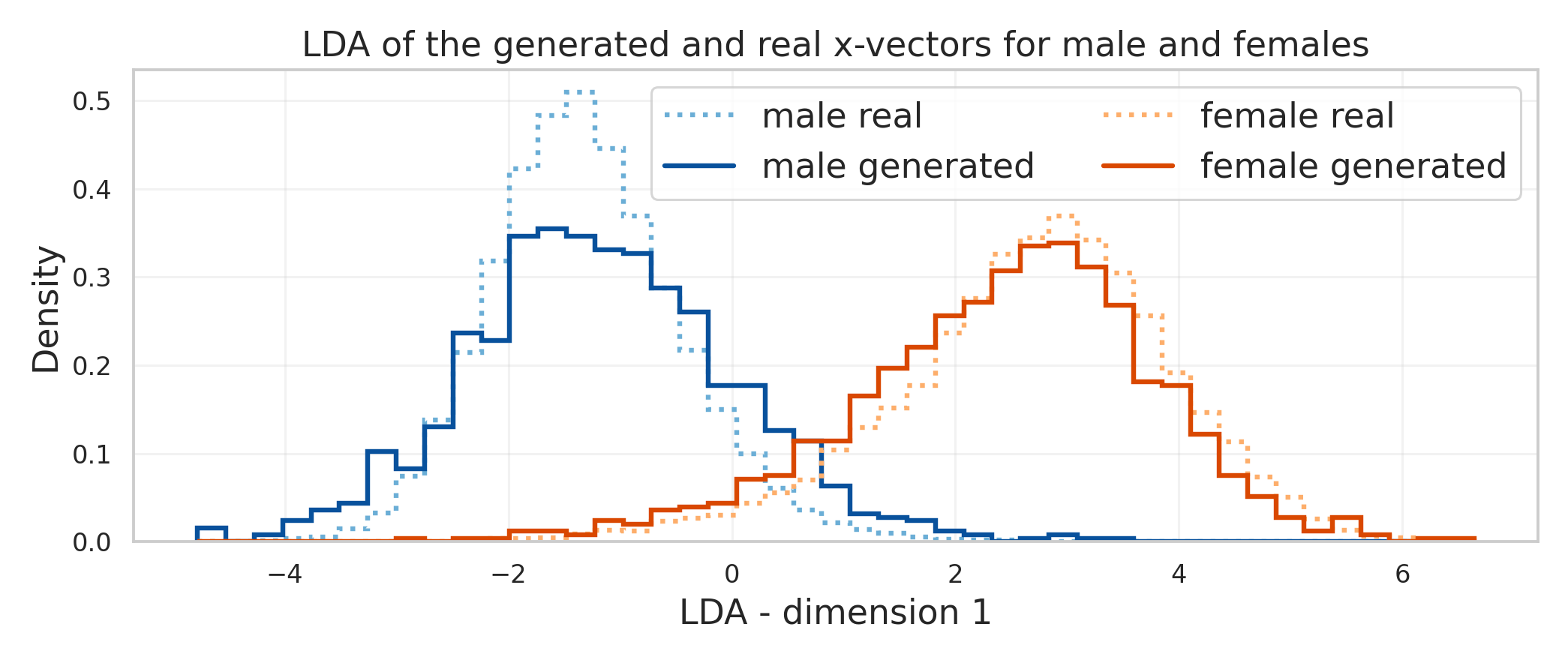}
    \caption{LDA visualization of generated and real x-vector distributions for males and females. The LDA is trained on the real dev-set x-vectors.}
    \label{fig:lda_gender}
    \vspace{-5mm}
\end{figure}

\begin{figure}[t]
    \vspace{-5mm}
    \centering
    \includegraphics[width=\linewidth]{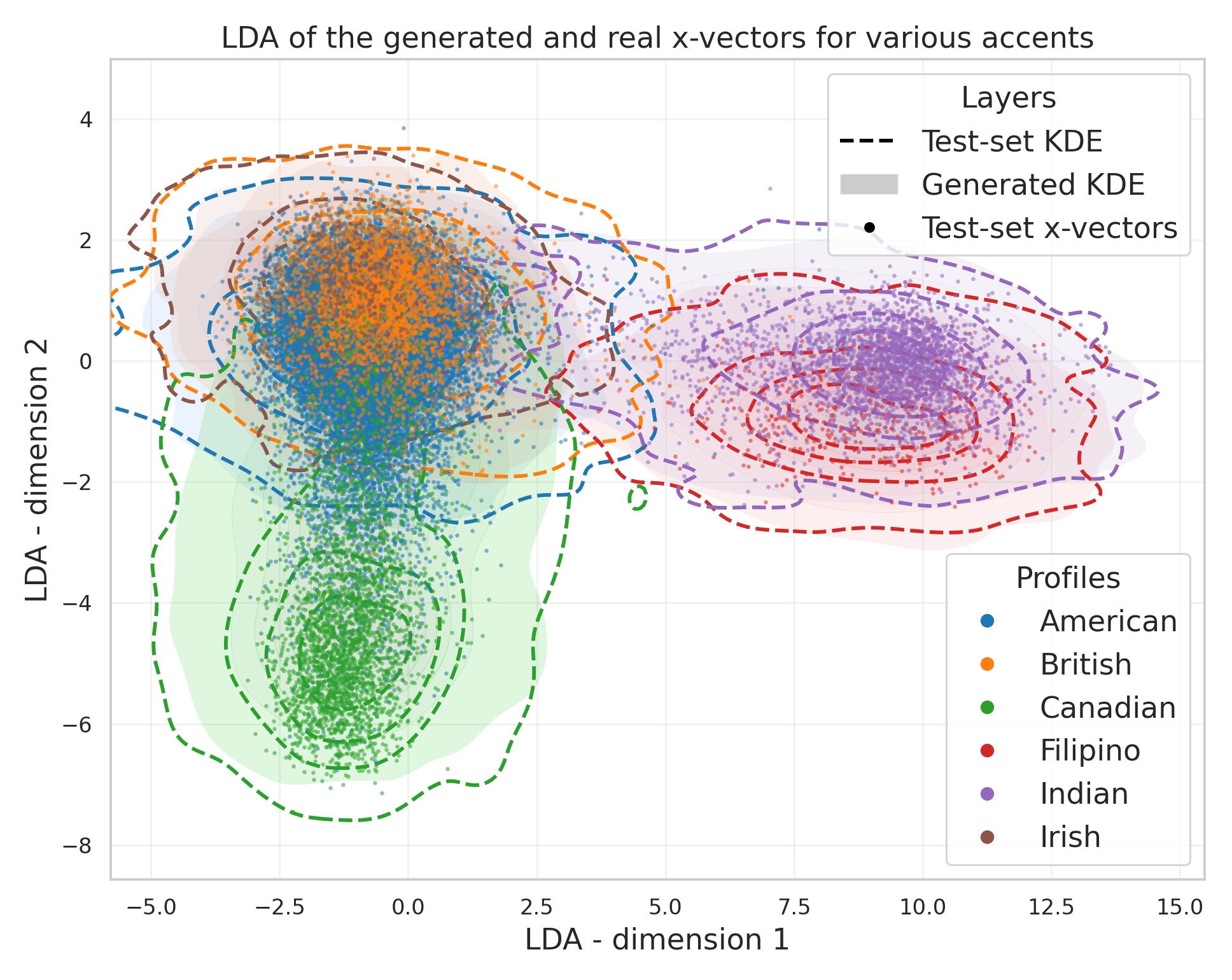}
    \caption{LDA visualization of generated and real x-vector distributions for the following accents: American, British, Canadian, Filipino, Indian, Irish. The LDA is trained on the real dev-set x-vectors.}
    \label{fig:lda_accent}
    \vspace{-7mm}
\end{figure}

Additionally, we provide LDA visualizations of the generated x-vectors compared to real x-vectors for various groups.
For a given characteristic (\texttt{accent=british}, for example), we provide a simplified description ("\texttt{a person with a british accent}") to the fine-tuned model, get the predicted GMM, and sample 10,000 generated x-vectors from it.
An LDA is trained on all the real x-vectors of the dev set corresponding to the set of categories selected, and the generated and real x-vectors from the test set are projected into this LDA.
Figure \ref{fig:lda_gender} shows the histogram of the male and female x-vectors on the LDA dimension, while
Figure \ref{fig:lda_accent} shows the distribution of the real and generated x-vectors for the American, Canadian, British, Filipino, Indian, Indian and Irish accents on the first 2 dimensions of the LDA.
A subset of accents has been selected, the ones with over 250 samples in the test set, to help with the readability.
Both figures show a clear separation of the different groups of x-vectors, as well as a close match between the generated distributions and the real x-vectors.

\begin{table}[b]
\vspace{-5mm}
\centering
\caption{Speaker-characteristic classification accuracy. Real accuracy measures performance on the real x-vectors of the test set, Generated accuracy measures performance on x-vectors sampled from generated GMMs.}
\label{tab:characteristic_accuracy}
\resizebox{\linewidth}{!}{%
\begin{tabular}{lcccc}
\toprule
Characteristic & Real & Generated & $\#$ classes & Classifier\\
\midrule
Gender & 99.1 & 98.4 & 2 & SVC\\
Age & 59.8 & 65.4 & 4 & SVR\\
Accent & 93.4 & 91.6 & 23 & SVC\\
Tone & 21.9 & 23.8 & 5 & SVR\\
Pitch & 37.5 & 28.2 & 7 & SVR\\
Pace & 37.4 & 31.6 & 5 & SVR \\
\bottomrule
\end{tabular}}
\end{table}

\subsection{Speaker-Characteristic Accuracy}

Table~\ref{tab:characteristic_accuracy} shows the classification accuracy obtained for each speaker characteristic on real and generated x-vectors. Overall, the generated samples preserve the requested characteristics well when those characteristics are strongly encoded in the x-vector space. For gender and accents, the SVM reaches high accuracies on real and generated x-vectors, showing that ProPS almost perfectly preserves the requested information. 
Age is also well preserved despite a lower accuracy. Interestingly, generated x-vectors obtain higher accuracy than real ones, suggesting that the generated age-conditioned GMMs may produce cleaner or more prototypical age distributions than the real held-out data, although age remains less separable than gender or accent.

The prosodic characteristics are more challenging. For the tone, both real and generated accuracies are much lower, close to chance for a five-class problem. This suggests that tone is poorly represented in the x-vector space or that the SVM is unable to reliably recover it. Pitch and pace show moderate real-test accuracies, but the generated accuracies drop by 9.3 and 5.8 absolute points, respectively. 
These gaps indicate that ProPS does not reliably preserve prosodic attributes, unlike demographic or accent-related attributes. 
This may reflect the fact that x-vectors are primarily optimized for speaker identity rather than detailed prosodic control, and that pitch and pace may be more sensitive to utterance-level variation than to stable speaker-profile structure. 
Overall, the results show that ProPS can generate controllable x-vectors for gender, accent, and age, but less reliable representations for finer-grained prosodic characteristics.

\section{Conclusion}

We presented ProPS, a framework for generating speaker-embedding distributions from natural language descriptions. ProPS maps free-form prompts to Gaussian mixture models in x-vector space using SBERT embeddings and a mixture density network initialized from profile-wise GMMs estimated on CapSpeech. Experiments show that the proposed approach provides a strong match to real x-vector distributions, with $K=16$ offering a practical trade-off between likelihood and profile coverage. The ablation study confirms the contribution of each training stage: GMM initialization anchors the model in real speaker space, pretraining associates generated prompts with profile components, and fine-tuning on human descriptions improves likelihood for realistic inputs.

Generated distributions preserve key speaker characteristics, especially gender, accent, and age, demonstrating that natural language can provide an effective control interface for speaker-profile synthesis in embedding space. In contrast, prosodic attributes such as pace, pitch, and tone are less reliably controlled, which is consistent with the fact that x-vectors are primarily designed to encode speaker identity rather than dynamic speaking style.

The method remains limited by the coverage and quality of CapSpeech profile labels and descriptions, and may inherit dataset biases or perform poorly for rare or ambiguous profiles. In addition, the current experiments are restricted to English data, limiting the range of usable descriptors and multilingual generalization. Future work will integrate ProPS-generated embeddings into downstream speech generation systems, evaluate perceptual quality and controllability after waveform generation, and explore richer multilingual speaker representations that better capture prosody and style.

\section*{Acknowledgment}

Generative AI tools were used during the preparation of this manuscript to assist with drafting, reformulation, and organization. All technical content, experimental design, benchmark implementation, results, and scientific claims were defined, verified, and edited by the authors.

This work was supported by the Office of the Director of National Intelligence (ODNI), Intelligence Advanced Research Projects Activity (IARPA), via the ARTS Program under contract D2023-2308110001. The views and conclusions contained herein are those of the authors and should not be interpreted as necessarily representing the official policies, either expressed or implied, of ODNI, IARPA, or the U.S. Government. The U.S. Government is authorized to reproduce and distribute reprints for governmental purposes notwithstanding any copyright annotation therein.

\bibliography{main} 
\bibliographystyle{IEEEtran}

\end{document}